\documentclass[superscriptaddress,showpacs,preprintnumbers,amsmath,amssymb]{revtex4}
\usepackage{graphicx} % Include figure files
\usepackage{dcolumn}  % Align table columns on decimal point
\usepackage{verbatim}
\usepackage{subfigure}
\usepackage{relsize} %% For the \babar command
\usepackage{calc} %% For entry evironment
\usepackage{xspace}
\usepackage{dcolumn} % Line up table entries by decimal points
\usepackage[amssymb,pstricks,Gray,squaren]{SIunits}

\graphicspath{{ps}}

\begin{document}

\title{
%\begin{flushright}{
%    \noindent
%    \hspace*{1.5in}Paper \#297\\
%    \hspace*{1.5in}Version 1.6\\ 
%    \hspace*{1.5in}July 14, 2009  }\\
%\end{flushright}
 \quad\\[0.5cm]  Evidence for $B\to K\eta'\gamma$ Decays at Belle
}

%%% Paper:    B -> K eta' gamma
%%% Journal:  Physical Review Letters
%%% Contacts: R. Wedd (r.wedd@pgrad.unimelb.edu.au)
%%% Non-responding authors or those who said NO are commented out.
%%% ====================================================================
%%% Click the RELOAD button on your web browser to see the updated file.
%%% ====================================================================
%%% Use \input{author} to insert this material into your latex file.
%%%%% Force institutions to appear in alphabetical order when typeset.
\affiliation{Budker Institute of Nuclear Physics, Novosibirsk}
\affiliation{Chiba University, Chiba}
\affiliation{University of Cincinnati, Cincinnati, Ohio 45221}
\affiliation{T. Ko\'{s}ciuszko Cracow University of Technology, Krakow}
%%%\affiliation{Department of Physics, Fu Jen Catholic University, Taipei}
%%%\affiliation{Justus-Liebig-Universit\"at Gie\ss{}en, Gie\ss{}en}
\affiliation{The Graduate University for Advanced Studies, Hayama}
%%%\affiliation{Gyeongsang National University, Chinju}
\affiliation{Hanyang University, Seoul}
\affiliation{University of Hawaii, Honolulu, Hawaii 96822}
\affiliation{High Energy Accelerator Research Organization (KEK), Tsukuba}
\affiliation{Hiroshima Institute of Technology, Hiroshima}
%%%\affiliation{University of Illinois at Urbana-Champaign, Urbana, Illinois 61801}
\affiliation{Institute of High Energy Physics, Chinese Academy of Sciences, Beijing}
%%%\affiliation{Institute of High Energy Physics, Vienna}
\affiliation{Institute of High Energy Physics, Protvino}
%%%\affiliation{Institute of Mathematical Sciences, Chennai}
%%%\affiliation{INFN - Sezione di Torino, Torino}
\affiliation{Institute for Theoretical and Experimental Physics, Moscow}
\affiliation{J. Stefan Institute, Ljubljana}
\affiliation{Kanagawa University, Yokohama}
\affiliation{Institut f\"ur Experimentelle Kernphysik, Universit\"at Karlsruhe, Karlsruhe}
\affiliation{Korea University, Seoul}
%%%\affiliation{Kyoto University, Kyoto}
\affiliation{Kyungpook National University, Taegu}
\affiliation{\'Ecole Polytechnique F\'ed\'erale de Lausanne (EPFL), Lausanne}
\affiliation{Faculty of Mathematics and Physics, University of Ljubljana, Ljubljana}
\affiliation{University of Maribor, Maribor}
\affiliation{University of Melbourne, School of Physics, Victoria 3010}
\affiliation{Nagoya University, Nagoya}
\affiliation{Nara Women's University, Nara}
\affiliation{National Central University, Chung-li}
\affiliation{National United University, Miao Li}
\affiliation{Department of Physics, National Taiwan University, Taipei}
\affiliation{H. Niewodniczanski Institute of Nuclear Physics, Krakow}
\affiliation{Nippon Dental University, Niigata}
\affiliation{Niigata University, Niigata}
\affiliation{University of Nova Gorica, Nova Gorica}
\affiliation{Novosibirsk State University, Novosibirsk}
\affiliation{Osaka City University, Osaka}
%%%\affiliation{Osaka University, Osaka}
\affiliation{Panjab University, Chandigarh}
%%%\affiliation{Peking University, Beijing}
%%%\affiliation{Princeton University, Princeton, New Jersey 08544}
%%%\affiliation{RIKEN BNL Research Center, Upton, New York 11973}
\affiliation{Saga University, Saga}
\affiliation{University of Science and Technology of China, Hefei}
\affiliation{Seoul National University, Seoul}
%%%\affiliation{Shinshu University, Nagano}
\affiliation{Sungkyunkwan University, Suwon}
\affiliation{University of Sydney, Sydney, New South Wales}
%%%\affiliation{Tata Institute of Fundamental Research, Mumbai}
%%%\affiliation{Toho University, Funabashi}
\affiliation{Tohoku Gakuin University, Tagajo}
\affiliation{Tohoku University, Sendai}
\affiliation{Department of Physics, University of Tokyo, Tokyo}
%%%\affiliation{Tokyo Institute of Technology, Tokyo}
\affiliation{Tokyo Metropolitan University, Tokyo}
\affiliation{Tokyo University of Agriculture and Technology, Tokyo}
%%%\affiliation{Toyama National College of Maritime Technology, Toyama}
\affiliation{IPNAS, Virginia Polytechnic Institute and State University, Blacksburg, Virginia 24061}
\affiliation{Yonsei University, Seoul}
  \author{R.~Wedd}\affiliation{University of Melbourne, School of Physics, Victoria 3010} % Melbourne
  \author{I.~Adachi}\affiliation{High Energy Accelerator Research Organization (KEK), Tsukuba} % KEK
  \author{H.~Aihara}\affiliation{Department of Physics, University of Tokyo, Tokyo} % Tokyo
  \author{K.~Arinstein}\affiliation{Budker Institute of Nuclear Physics, Novosibirsk}\affiliation{Novosibirsk State University, Novosibirsk} % BINP
% \author{T.~Aso}\affiliation{Toyama National College of Maritime Technology, Toyama} % Toyama
  \author{V.~Aulchenko}\affiliation{Budker Institute of Nuclear Physics, Novosibirsk}\affiliation{Novosibirsk State University, Novosibirsk} % BINP
% \author{T.~Aushev}\affiliation{\'Ecole Polytechnique F\'ed\'erale de Lausanne (EPFL), Lausanne}\affiliation{Institute for Theoretical and Experimental Physics, Moscow} % ITEP
% \author{T.~Aziz}\affiliation{Tata Institute of Fundamental Research, Mumbai} % Tata
% \author{S.~Bahinipati}\affiliation{University of Cincinnati, Cincinnati, Ohio 45221} % Cincinnati
  \author{A.~M.~Bakich}\affiliation{University of Sydney, Sydney, New South Wales} % Sydney
  \author{V.~Balagura}\affiliation{Institute for Theoretical and Experimental Physics, Moscow} % ITEP
% \author{Y.~Ban}\affiliation{Peking University, Beijing} % Peking
  \author{E.~Barberio}\affiliation{University of Melbourne, School of Physics, Victoria 3010} % Melbourne
  \author{A.~Bay}\affiliation{\'Ecole Polytechnique F\'ed\'erale de Lausanne (EPFL), Lausanne} % Lausanne
% \author{I.~Bedny}\affiliation{Budker Institute of Nuclear Physics, Novosibirsk}\affiliation{Novosibirsk State University, Novosibirsk} % BINP
  \author{K.~Belous}\affiliation{Institute of High Energy Physics, Protvino} % Protvino
% ...
% BINP
\collaboration{The Belle Collaboration}

\begin{abstract}

We present the results of a search for the radiative decay $B \rightarrow K \eta' \gamma$ and find 
evidence for $B^{+} \rightarrow K^{+} \eta' \gamma$ decays at the $3.3$ standard deviation level with 
a partial branching fraction of $(3.6\pm1.2\pm0.4)\times10^{-6}$, where the first error is statistical 
and the second systematic.  This measurement is restricted to the region of combined $K\eta'$
invariant mass less than $3.4\,\giga\electronvolt\!/c^2$.
A 90\% confidence level upper limit of $6.4\times10^{-6}$ is obtained for the decay 
$B^{0} \rightarrow K^{0} \eta' \gamma$ in the same $K\eta'$ invariant mass region. These results are 
obtained from a 605\,$\femto\reciprocal\barn\xspace$ 
data sample containing 657 $\times 10^6 B\overline{B}$ pairs collected at the $\Upsilon(4S)$ resonance
with the Belle detector at the KEKB asymmetric-energy $e^+ e^-$ collider.

\end{abstract}


\maketitle

%%%% >>>> keep the final version single-spaced
\tighten

{\renewcommand{\thefootnote}{\fnsymbol{footnote}}}
\setcounter{footnote}{0}

%Radiative $B$ meson decays play an important role in indirect searches for physics beyond the Standard Model (SM). 
%They 
Radiative $B$ meson decays proceed primarily through the flavour changing neutral current (FCNC) quark-level process $ b \rightarrow s \gamma$.
%, the former being the most common transition. 
FCNC processes are forbidden at tree-level within the Standard Model (SM), and hence $ b \rightarrow s \gamma$ must proceed via radiative loop diagrams. 
As loop processes may include unknown heavy particles mediating the loop, any disparity between experimental measurement and 
SM prediction could be evidence of such new particles.

The world average experimental branching fraction (BF) for the meson-level process 
$B \rightarrow X_{s} \gamma$ ($(3.55\pm0.26)\times10^{-4}$ \cite{exp bf}) 
and the theoretical SM predictions ($(3.15\pm0.23)\times10^{-4}$ \cite{th bf}) are consistent. 
Measurements of individual exclusive $B \rightarrow X_{s} \gamma$ modes, such as $B \rightarrow K \eta' \gamma$, provide consistency checks on the agreement 
between theory and experiment, and improve our understanding of the hadronization process in $B \rightarrow X_{s} \gamma$ and $B \rightarrow X_{s} l^{+} l^{-}$.
%A charge-parity ($CP$) violation study will also be 
%possible if sufficient statistics are accumulated for the decay $B^{0} \rightarrow K^{0} \eta' \gamma$.

%A similar 3-body decay mode $B \rightarrow K \eta \gamma$ has already been measured at Belle \cite{nishida} 
%and by the {\sc BaBar} collaboration \cite{babar}.
%, with average BFs of ${\cal B}(\BToKetagch) = (9.4\pm1.1)\times10^{-5}$ and ${\cal B}(\BToKetagn) = (1.1\pm0.2)\times10^{-5}$ \cite{PDG08}. 
%A comparison between the branching fractions of this mode 
%and $B \rightarrow K \eta' \gamma$ will be of interest. 
The analysis of the decay mode $B \rightarrow K \eta' \gamma$ uses $605\,\femto\reciprocal\barn\xspace$ of data collected 
at the $\Upsilon(4S)$ resonance with the Belle detector. This decay was previously studied by {\sc BaBar} 
who set upper limits (ULs) of ${\cal B}(B^{+} \rightarrow K^{+} \eta' \gamma) < 4.2\times 10^{-6}$ and ${\cal B}(B^{0} \rightarrow K^{0} \eta' \gamma) < 6.6\times 10^{-6}$ at 90\% confidence level (CL) from an analysis of 
$211\,\femto\reciprocal\barn\xspace$ of data \cite{babar}. 

A previous comparison of the modes $B \rightarrow K \eta$ and $B \rightarrow K \eta'$ showed a surprising 
suppression of the former with respect to the latter, due to destructive interference between two penguin 
amplitudes \cite{Lipkin}. The ability of QCD factorization techniques to 
correctly predict this BF hierarchy has been demonstrated, though the errors are large \cite{neubert}. A similar 
comparison of the observed $B \rightarrow K \eta \gamma$  mode \cite{nishida}\cite{babar} and the previous upper limits for $B \rightarrow K \eta' \gamma$ 
shows the opposite trend for these BF's. The analogous QCD calculation has not yet been performed for these decay modes. 
Measurement 
of $B \rightarrow K \eta' \gamma$ is an important test of such a calculation.  
In addition, a time-dependent charge-parity ($CP$) asymmetry measurement will be possible if there are sufficient 
statistics for the decay $B^{0} \rightarrow K^{0}_{S} \eta' \gamma$ \cite{CC}. Such mixing-induced CP 
asymmetries are suppressed within the SM, but some beyond-SM theories involving right-handed currents allow 
them to be large, even when the $B^{0} \rightarrow K^{0}_{S} \eta' \gamma$ BF agrees with SM predictions \cite{TDPCV}.

The Belle detector is designed to identify and measure 
particles from $\Upsilon(4S)\rightarrow B {\kern 0.18em\overline{\kern -0.18em B}}$ decays \cite{Belle}. It is located at the 
interaction point (IP) of the KEKB accelerator in Tsukuba, Japan, which collides electrons and 
positrons at energies of $8\,\giga\electronvolt$ and $3.5\,\giga\electronvolt$, respectively \cite{KEKB}. Charged particle tracking 
and momentum measurements are provided by a
% multi-layered 
silicon vertex detector (SVD) and a 
%50 layer 
helium/ethane central drift chamber (CDC). Particle 
identification (PID) is performed using the CDC in conjunction with an array of aerogel Cherenkov counters (ACC) and 
time-of-flight scintillators (TOF). Electrons and photons are identified and their energy measured by an 
electromagnetic calorimeter (ECL) of CsI(Tl) crystals. These detectors are within a 1.5T magnetic field provided by a 
superconducting solenoid. A layered iron and resistive plate counter detector outside the solenoid 
gives $K_{L}/\mu$ separation.

The primary signature of a $B \rightarrow K \eta' \gamma$ event is a high energy photon, which we reconstruct as an isolated shower in 
the ECL barrel ($32^{\circ} < \theta < 129^{\circ}$), with shape consistent with a single photon hypothesis and no associated charged track. 
Only photons within the range 
$1.8\,\giga\electronvolt < E_{\gamma}^{*} < 3.4\,\giga\electronvolt$ are considered, where the asterisk denotes the $e^{+}e^{-}$ center-of-mass (CM) 
frame. All other photons in the analysis are required to have energies greater than $50\,\mega\electronvolt$. 
Contamination from $\eta$ and $\pi^{0}$ mesons decaying to $\gamma\gamma$ is reduced using a likelihood technique 
based on probability density functions (PDFs) from Monte Carlo (MC) simulated data. 
If more than one photon in an event passes the selection, the one with the highest energy is taken.

\begin{figure}[t!]
\small
\centering
\subfigure[$M_{\rm bc}$ in $\Delta E$ signal region for $B^{+} \rightarrow K^{+} \eta' \gamma$]{ \includegraphics[width=0.35\columnwidth,height=0.19\linewidth]{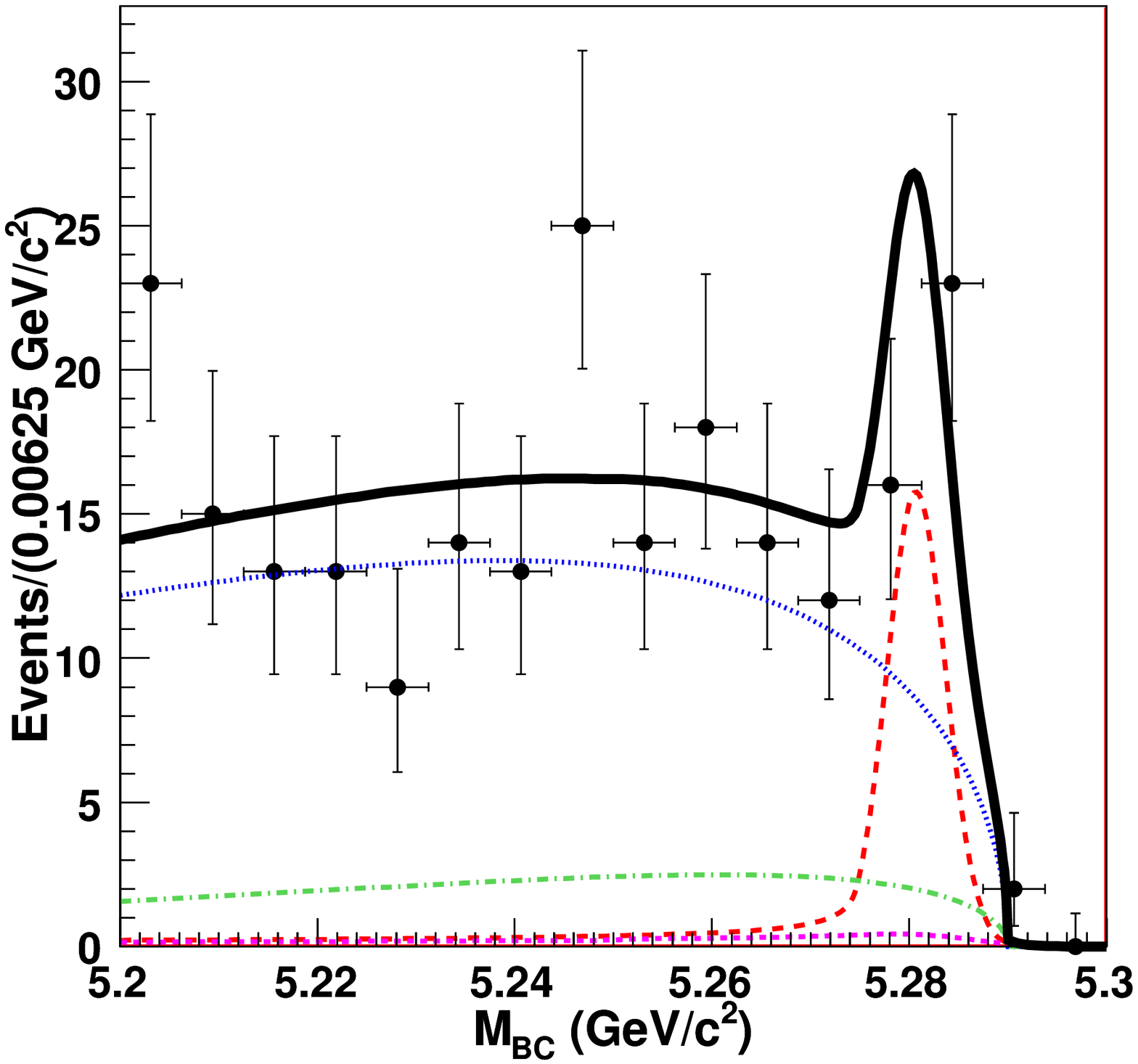}\hspace{1.0cm}} 
\subfigure[$\Delta E$ in $M_{\rm bc}$ signal region for $B^{+} \rightarrow K^{+} \eta' \gamma$]{ \includegraphics[width=0.35\columnwidth,height=0.19\linewidth]{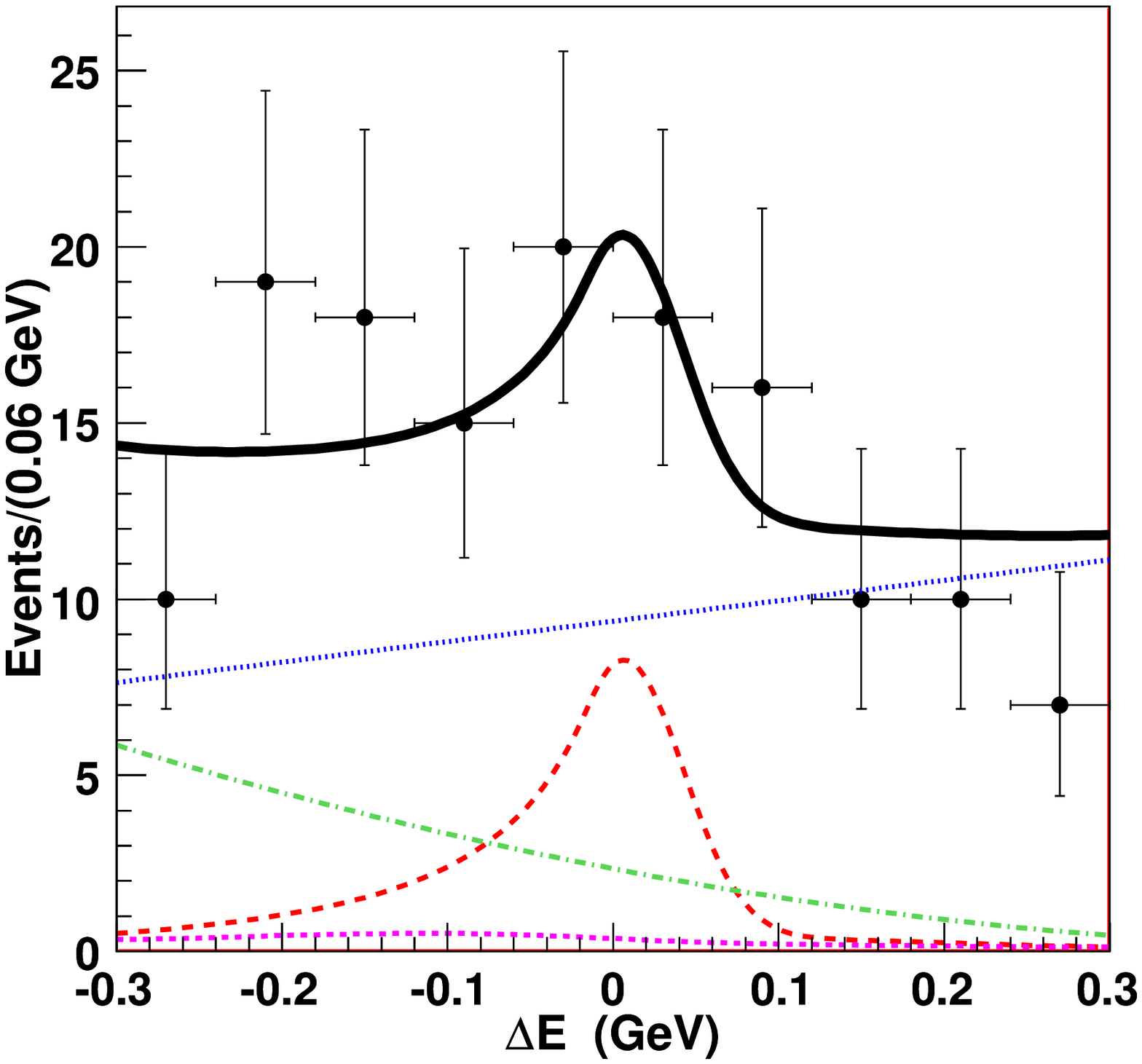}\hspace{1.0cm}} 
\subfigure[$M_{\rm bc}$ in $\Delta E$ signal region for $B^{0} \rightarrow K^{0}_{S} \eta' \gamma$]{ \includegraphics[width=0.35\columnwidth,height=0.19\linewidth]{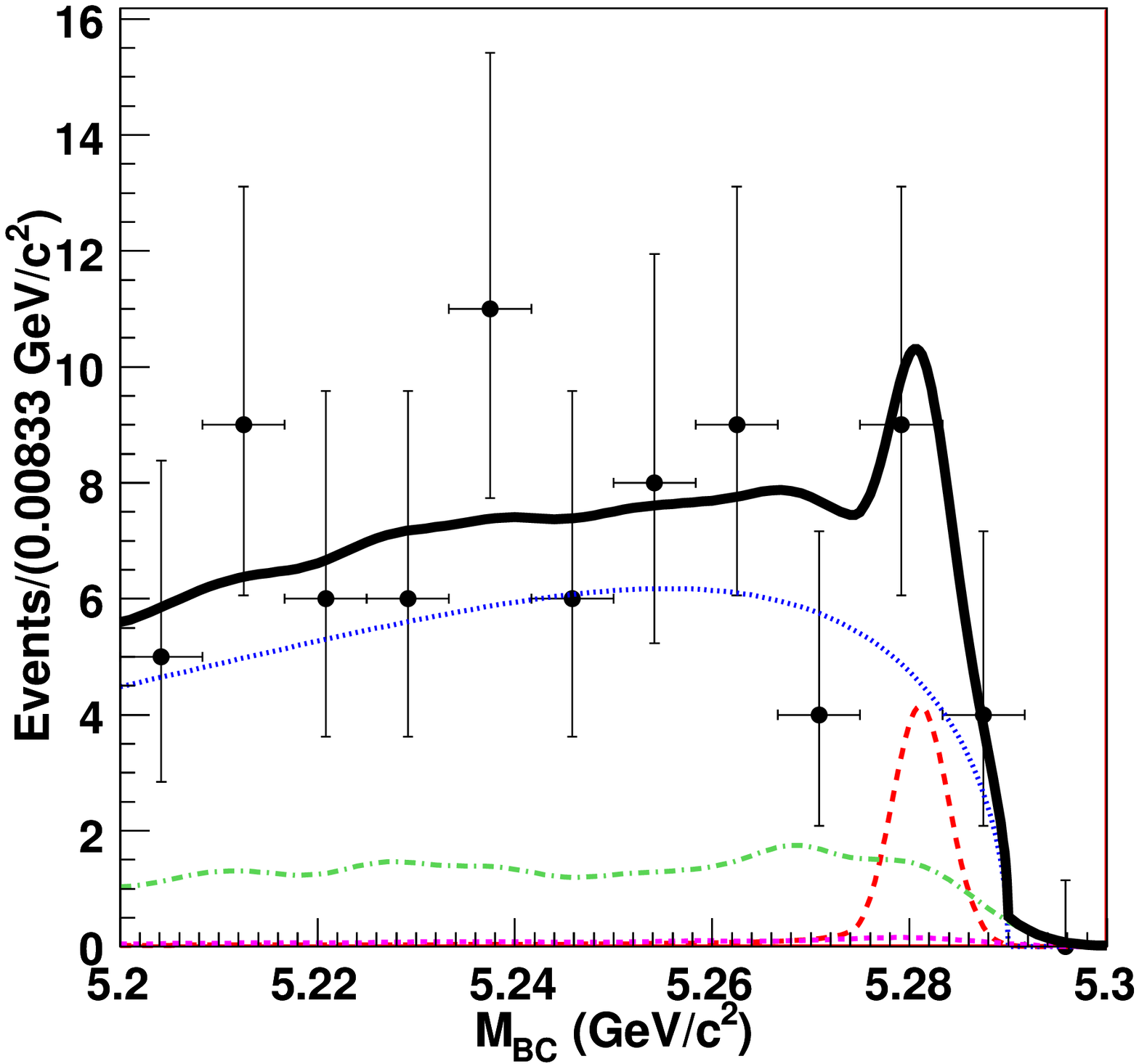}\hspace{1.0cm}} 
\subfigure[$\Delta E$ in $M_{\rm bc}$ signal region for $B^{0} \rightarrow K^{0}_{S} \eta' \gamma$]{ \includegraphics[width=0.35\columnwidth,height=0.19\linewidth]{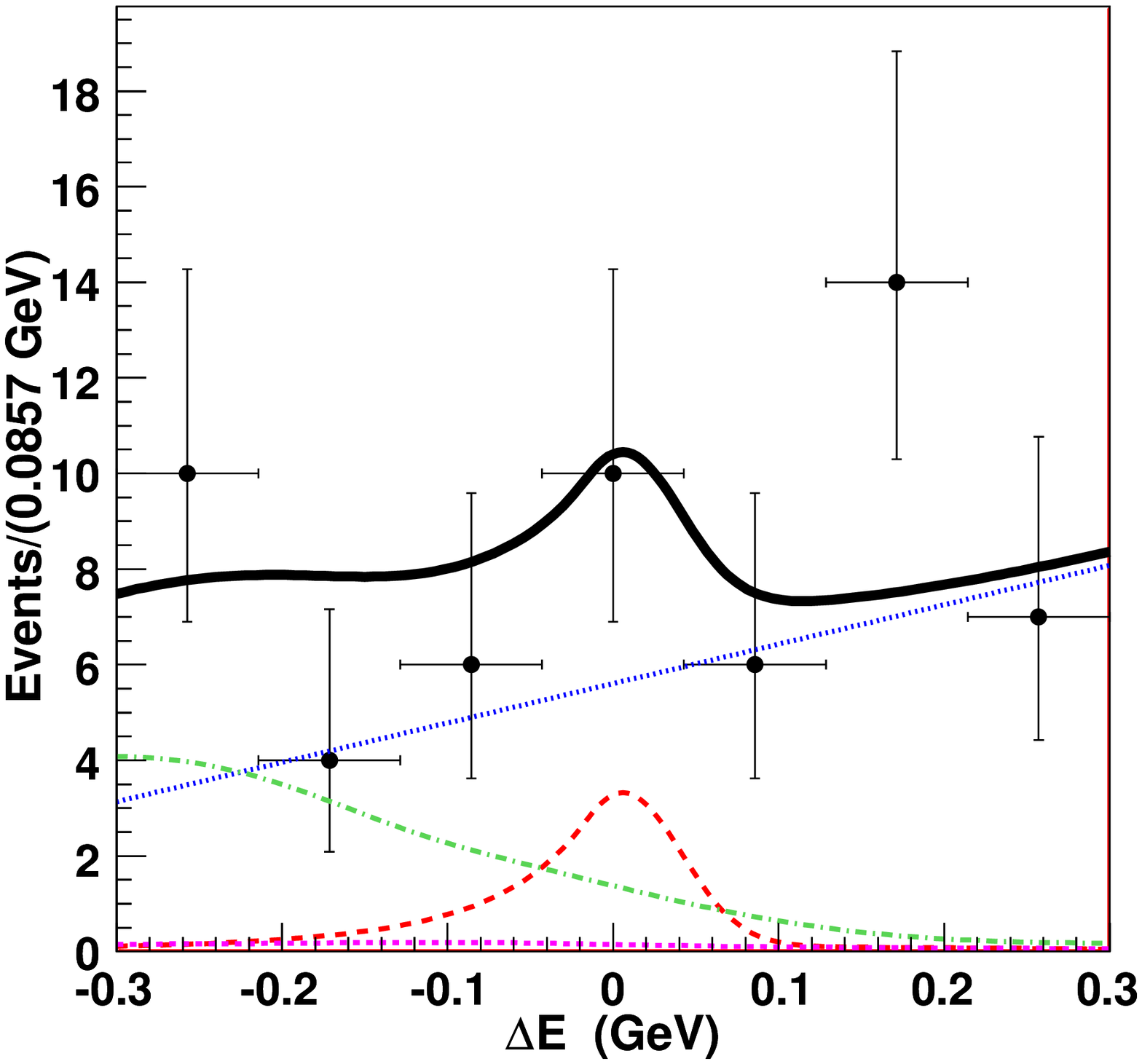}\hspace{1.0cm}} 
\vskip -.15cm
\caption[]{Projections to the signal region from the fits to $605\femto\reciprocal\barn\xspace$ of data. The solid lines are the combined background plus $K \eta' \gamma$ functions,
the dotted lines are the $e^{+}e^{-}\rightarrow q \overline q$ functions, the dot-dashed the $b \rightarrow c$ functions, the long dashed the $K \eta' \gamma$ functions and the short dashed the $b \rightarrow u,d,s$ functions. \label{fits} }
%\vskip -0.8cm
\end{figure}

Candidate $\eta'$ mesons are reconstructed in the $\eta' \rightarrow \eta  \pi^{+} \pi^{-}$ and $\eta' \rightarrow \rho^{0} \gamma$ modes. 
Here, $\eta$ mesons are 
reconstructed as $\eta \rightarrow \gamma \gamma$
and $\eta \rightarrow \pi^{+} \pi^{-} \pi^{0}$, $\rho^{0}$ and $K^{0}_{S}$ mesons are reconstructed as $\pi^{+}\pi^{-}$ pairs, and $\pi^{0}$ mesons as $\gamma\gamma$ pairs. 
%All charged tracks except for $K^{0}_{S}$ daughter pions
%are required to pass within $1.5\,{\mathrm{cm}}$ of the beam axis, within $4\,{\mathrm{cm}}$ of the $e^{+}e^{-}$ collision point 
%along the beam axis, and have momentum greater than $100\,\mega\electronvolt c$. 
Charged kaons and pions are separated using a PID selection with an efficiency for kaons (pions) of 85\% (98\%) and a
misidentification probability of 12\% (14\%).
%$dE/dx$ information from the CDC, light yield in the ACC, 
%and timing information from the TOF. 
PID information combined with a measurement of the energy deposited in the ECL is used 
to reduce electron contamination.

$K^{0}_{S}$ candidates must pass a set of momentum-dependent selection criteria based on proximity to the IP, flight length, and the 
angle between the momentum vector and reconstructed vertex vector. Candidates with invariant masses $10\,\mega\electronvolt c^{2}$ 
or more from the nominal $K^{0}_{S}$ mass of $497.7\,\mega\electronvolt c^{2}$ \cite{PDG} are rejected, where the allowed range is 
equivalent to a window of 4 standard deviations ($\sigma$) of the natural mass width convolved with detector mass resolution. 
Neutral pion candidates must have momenta greater than $100\,\mega\electronvolt c$ and $\gamma\gamma$ combined invariant 
masses in the range $119\,\mega\electronvolt c^{2}\,(2.5\sigma) < M_{\gamma\gamma} < 152\,\mega\electronvolt c^{2}\,(3\sigma)$.

$\eta \rightarrow \gamma \gamma$ candidates are reconstructed using photons of energy greater than $100\,\mega\electronvolt$, and must have invariant 
masses in the range $490\,\mega\electronvolt c^{2}(4\sigma)<M_{\gamma\gamma}<590\,\mega\electronvolt c^{2}(3\sigma)$. They are also 
required to satisfy the helicity angle requirement $|\cos\theta_{\mathrm{hel}}|<0.9 $, where
$\theta_{\mathrm{hel}}$\, is the angle between the momentum vectors of the $\eta'$ and one of the decay $\gamma$'s in the 
$\eta$ rest frame. $\eta \rightarrow \pi^{+} \pi^{-} \pi^{0}$ candidates must have invariant masses in the range 
$536\,\mega\electronvolt c^{2}\,(3\sigma)<M_{\pi^{+}\pi^{-}\pi^{0}}<560\,\mega\electronvolt c^{2}\,(3\sigma)$. The momenta
of all $\eta$ candidates are then corrected using a mass-constrained fit. $\rho^{0}$ candidates must have invariant masses within 
the range $550\,\mega\electronvolt c^{2}\,(3\sigma)<M_{\pi^{+}\pi^{-}}<950\,\mega\electronvolt c^{2}\,(3\sigma)$ and also pass a helicity requirement similar to
 the $\eta$.
%: \etahelcut~.
%, where here
%\etaHel~ is the angle between the momentum vectors of the \etap and one of the decay $\pi$'s in the $\rho^{0}$ frame. 
A fit constraining the $\pi^{+}\pi^{-}$ vectors to a common vertex must succeed.

$\eta' \rightarrow \rho^{0} \gamma$ candidates must have invariant masses in the range 
$945~\,\mega\electronvolt c^{2}\,(2\sigma)<M_{\rho^{0}\gamma}<970~\,\mega\electronvolt c^{2}\,(2\sigma)$, and the photon
must have energy greater than $200\,\mega\electronvolt$. $\eta' \rightarrow \eta  \pi^{+} \pi^{-}$ candidates must have invariant masses within the range
$950~\,\mega\electronvolt c^{2}\,(2\sigma)<M_{\eta\pi^+\pi^-}<965~\,\mega\electronvolt c^{2}\,(2\sigma)$. 
Both types of candidates are required to have momenta greater than $1.0\,\giga\electronvolt c$ as measured in the CM frame.

The invariant mass of the $K\eta'$  system ($M_{K\eta'}$) is required to be less than $3.4\,\giga\electronvolt c^{2}$. This requirement 
removes some background while retaining most of the expected $M_{K\eta'}$ spectrum.
%The signal photon is added to a $K\etap$ to form a $B$ candidate. 
Two kinematic variables are defined: 
$M_{\rm bc} \equiv (1/c^{2})\sqrt{E_{\mathrm{beam}}^{2} - (\vec{p}_{K\eta'}^{*} + \vec{p}_{\gamma}^{*})^{2}c^{2} }$ and 
$\Delta E \equiv E^{*}_{B} - E_{\mathrm{beam}}$, 
where $E^{*}_{B}$ is the energy of the candidate $B$ meson in the CM frame, $E_{\mathrm{beam}}$ is half the total CM energy ($\sqrt{s}/2$), and 
$\vec{p}_{K\eta'}^{*}$ and $\vec{p}_{\gamma}^{*}$ are the CM frame momenta of the $K\eta'$ combination and
the signal photon, respectively. In the calculation of $M_{\rm bc}$, the momentum of the signal photon is rescaled to be  
$p^{*}_{\gamma} = (1/c)(E_{\mathrm{beam}}  - E^{*} _{K\eta'})$. $B$ candidates must satisfy $|\Delta E| < 0.3~\giga\electronvolt$ and $5.20~\giga\electronvolt c^{2} < M_{\rm bc} < 5.29~\giga\electronvolt c^{2}$, and the signal region is defined 
as $-0.1~\giga\electronvolt < \Delta E < 0.07~\giga\electronvolt$ and $5.27~\giga\electronvolt c^{2} < M_{\rm bc} < 5.29~\giga\electronvolt c^{2}$.

Backgrounds to $K \eta' \gamma$ events are estimated using large MC samples \cite{MC1}. The dominant background is from $e^{+}e^{-}\rightarrow q \overline q$ 
($q = u,d,s,c$) continuum processes. To reduce this we form a Fisher discriminant \cite{fisher} 
from 16 modified Fox-Wolfram moments \cite{fw} and the scalar sum of the event transverse momenta. A  likelihood ratio (LR) is 
formed from the optimized Fisher discriminant, 
the cosine of the angle between the $B$ meson flight direction and the positron beam axis, and the distance 
along the positron beam axis between the two $B$ meson vertices, which are reconstructed from the SVD and CDC response to charged 
particles. The signal regions of the LR distributions are chosen 
%in 7 bins of the $B$-flavour tagging variable $qr$ 
by maximizing the figure of merit defined as 
${\cal N}^{}_{S}/(\sqrt{{\cal N}^{}_{S}+{\cal N}^{}_{SB}})$, where ${\cal N}^{}_{S}$ is the number of $B \rightarrow K \eta' \gamma$ MC events that lie above a certain LR value and ${\cal N}^{}_{SB}$ 
is the corresponding number of sideband data events lying above the same value. The data sideband regions are defined as $M_{\rm bc}<5.26\,\giga\electronvolt\!/c^2$
and either $\Delta E<-0.2\,\giga\electronvolt$ or $\Delta E>0.1\,\giga\electronvolt$. The BF central values found in the 
{\sc BaBar} $K \eta' \gamma$ analysis \cite{babar} 
are used to calculate ${\cal N}^{}_{S}$, and the sideband data is scaled by the ratio of $q \overline q$ MC events in the full fitting region to 
those in the sideband region.
%Both samples are scaled to the number of events expected in 
%$605\,\femto\reciprocal\barn\xspace$ of data, using the central values of the BFs found in the {\sc BaBar} $K \eta' \gamma$ analysis as hypothetical signal BFs, 
%and the ratio of events in the full fitting region to the sideband region in $q \overline q$ MC as the data sideband 
%scaling factor. 
To maximize discrimination, the optimization of the figure of merit is performed in bins of flavour-tagging quality, 
as calculated by the Belle $B$-flavour tagging algorithm \cite{tag_algorithm}. 
%reconstructs a second $B$ meson from the remaining event particles 
%and calculates two parameters: the most likely flavour of this second $B$ meson, $q$ 
%($q=+1$ for \Bp,\Bz; $q=-1$ for \Bm,\Bzbar), and the flavour assignment quality, $r$ (ranging from 0 for no flavour 
%information to 1 for unambiguous flavour assignment). 
The LR requirements are 38\% efficient for signal MC and remove 98\% of background. 

A significant proportion of the MC events modelled as $b \rightarrow c$ processes that pass the above selection criteria 
are found to include a $D^{0}$ meson. 
%These events are suppressed by successively combining the charged kaon candidate with each of the charged pions in the event. 
Any kaon candidate which forms an invariant mass within the range $1.84\,\giga\electronvolt\!/c^2<M_{ K^{-} \pi^{+}}<1.89\,\giga\electronvolt\!/c^2$ 
when combined with any charged pion in the same event is removed from consideration, in order to suppress $D^{0}\rightarrow K^{-}\pi^{+}$ decays. 
%In addition, 
$B \rightarrow J/\psi K \rightarrow(\eta'\gamma)K$ events are suppressed by vetoing candidates with a combined 
$\eta'\gamma$ invariant mass within $\pm25\mega\electronvolt c^{2}$ of the nominal $J/\psi$ mass \cite{PDG}. This reduces this 
background to a negligible level.

%More than one $B$ candidate can be reconstructed from each event. 
On average, 1.24 candidates per signal MC event pass the selection criteria. 
%Because of the high particle multiplicity of the reconstructed modes, a 
A series of selection criteria are used to choose the best candidate, including the lowest 
$B$ vertex $\chi^{2}$, the lowest $\rho^{0}\rightarrow\pi^{+}\pi^{-}$ or $\eta' \rightarrow \eta  \pi^{+} \pi^{-}$ vertex $\chi^{2}$, the reconstruction 
with $\eta$ candidate invariant mass closest to the nominal value \cite{PDG}, the highest $E_{\gamma}$ from $\eta' \rightarrow \rho^{0} \gamma$,
and the lowest $K^{0}_{S}$ vertex $\chi^{2}$. This technique selects the correct candidate in $76\%$ of cases.

%When the $B$ candidates differ in included charged particles, the best candidate with the lowest 
%$B$ vertex $\chi^{2}$ is chosen. This differentiates between the majority of candidates, as the most commonly mis-reconstructed
%particle in both charged and neutral $K \eta' \gamma$ modes is the kaon. When each candidate includes the same charged particles in 
%different configurations, the lowest $\rho^{0}\rightarrow\pi^{+}\pi^{-}$ or $\eta' \rightarrow \eta  \pi^{+} \pi^{-}$ vertex $\chi^{2}$ is used to chose the best candidate. 
%When the $B$ candidates differ in the neutral components and the final states include an $\eta$, the reconstruction 
%with the $\eta$ candidate invariant mass closest to the nominal value \cite{PDG} is taken. When the final states include 
%a \rho^{0}, the candidate with the highest $E_{\gamma}$ from $\eta' \rightarrow \rho^{0} \gamma$ is taken.

Signal yields are extracted using extended unbinned maximum likelihood fits to $\Delta E$ and $M_{\rm bc}$. All 
reconstructed charged final states are combined into one fitted distribution, and all reconstructed neutral 
final states are combined into another. PDF parameters 
are determined from MC distributions. The $K \eta' \gamma$ distribution is modelled with a Crystal Ball line shape (CBLS)
\cite{cbls} function for $M_{\rm bc}$ and CBLS plus Gaussian functions with common means and relative widths for $\Delta E$. 
The $b \rightarrow c$ background is modelled with an
ARGUS \cite{argus} function for $M_{\rm bc}$ and a second order Chebyshev polynomial for $\Delta E$. 
%The shape parameter for the ARGUS function is dependent on $\Delta E$ and the function is normalized using the condition 
%$\int F(x|y)dx \equiv 1$ for all values of $y$, where $x=$M_{\rm bc}$$ and $y=$\Delta E$$. 
%Correlations found between $M_{\rm bc}$ and $\Delta E$ in $b \rightarrow c$ MC are 
%included in the fit by the use of a conditional function. 
The $b \rightarrow u,d,s$ background is modelled with a 2D Keys PDF \cite{keys}.
%, which is a combination of Gaussians centered at each $b \rightarrow u,d,s$ MC point. The width of each Gaussian is dependent on the surrounding MC density. The shape of the 
%$M_{\rm bc}$ and $\Delta E$ distributions for $b \rightarrow u,d,s$ MC are found to be independent of the LR cuts; these cuts are relaxed 
%to increase MC statistics and reduce parameter uncertainty. 
The $e^{+}e^{-}\rightarrow q \overline q$ distribution is modelled 
with an ARGUS function for $M_{\rm bc}$ and a first order Chebyshev polynomial for $\Delta E$.

The means and widths of the CBLS functions describing the $K \eta' \gamma$ distribution in $\Delta E$ and $M_{\rm bc}$ are calibrated 
using large control samples of $B \rightarrow K^{*}(892) \gamma$ data and MC. The PDFs of the $K \eta' \gamma$, $e^{+}e^{-}\rightarrow q \overline q$, $b \rightarrow c$ and $b \rightarrow u,d,s$ distributions
are combined and used to fit the $605\,\femto\reciprocal\barn\xspace$ of accumulated data. The normalizations of the $b \rightarrow c$ and $b \rightarrow u,d,s$ components are
fixed to values expected from MC studies; the $K \eta' \gamma$ and $e^{+}e^{-}\rightarrow q \overline q$ normalizations and the $e^{+}e^{-}\rightarrow q \overline q$ function parameters 
are allowed to float, except for the ARGUS endpoint, which is fixed to $5.29\,\giga\electronvolt\!/c^2$. 
Figure \ref{fits} shows the results of the $B^{+} \rightarrow K^{+} \eta' \gamma$ and 
$B^{0} \rightarrow K^{0}_{S} \eta' \gamma$ fits to data, where the $M_{\rm bc}$ plots show a projection in the $\Delta E$ signal 
region and the $\Delta E$ plots show a projection in the $M_{\rm bc}$ signal region. 

Table \ref{bf} shows the measured yields and signal significances for the fits to data. We find $33^{+12}_{-11}$ 
events for the fit to the charged modes and 
$5^{+5}_{-4}$ events for the fit to the neutral modes.
%, with statistical significances of $3.4\sigma$ and $1.3\sigma$, respectively. 
The signal significance is defined as $\sqrt{-2{\mathrm{ln}}(L_{0}/L_{\mathrm{max}})}$, 
where $L_{\mathrm{max}}$ and $L_{0}$ 
are the values of the likelihood function when the signal yield is floated or fixed to zero, respectively. 
The systematic errors described below are included in the significances by convolving the likelihood functions 
with Gaussians of width defined by the magnitude of the errors. The signal significances including 
systematic errors are $3.3\sigma$ and $1.3\sigma$ for the charged modes and neutral modes, respectively.

\begin{table*}[t!]
\setlength{\extrarowheight}{3.5pt}
\begin{center}
\caption{The yields, efficiencies ($\varepsilon$), daughter branching fraction products ($\prod$), measured branching fractions (${\cal B}$), signal significances including systematics (${\cal S}$) and 90\% CL ULs for the measured decays.\label{bf}}
\vspace{0.3cm}
\begin{tabular}{ c  c  c  c  c  c  c  c  c  c  c  c  c  c  c  c  c  c  c}\hline\hline
{Mode} &&&  Yield(events) &&&$\varepsilon$   &&& $\prod$ &&& ${\cal B}$($10^{-6}$) &&& {$\cal S$}($\sigma$)  &&& UL($10^{-6}$) \\\hline
\multicolumn{1}{ c }{ $B^{+} \rightarrow K^{+} \eta' \gamma$ }  &&&  $32.6^{+11.8}_{-10.8}$&&& 0.024 &&& 0.571   &&& $3.6\pm1.2\pm0.4$ &&& 3.3 &&& 5.6 \\ %&&& 3.43  and 4.8 without sys (5.0 with)\\
\multicolumn{1}{ c }{ $B^{0} \rightarrow K^{0} \eta' \gamma$ }   &&&  $5.1^{+5.0}_{-4.0}$   &&& 0.016 &&& 0.197   &&& $2.5^{+2.4 +0.4}_{-1.9 -0.5}$ &&& 1.3 &&& 6.4 \\\hline\hline %&& 1.33 and 6.2 without sys\\\hline
\end{tabular}
\end{center}
\vskip -0.8cm
\end{table*}

Any bias in the fitting process is determined using pseudoexperiments of MC scaled to the yields returned 
by the fits to data. The $e^{+}e^{-}\rightarrow q \overline q$ and $b \rightarrow c$ MC components are generated from the shape of the PDFs and 
combined with fully simulated MC for the $K \eta' \gamma$ and $b \rightarrow u,d,s$ components. The analysis of 1000 pseudoexperiments
finds no significant bias.

The signal MC reconstruction efficiencies are calculated as the number of $K \eta' \gamma$ events passing the
selection criteria divided by the number generated. To calibrate for $M_{K\eta'}$ dependence, the 
efficiencies in 10 bins across $M_{K\eta'}$ are weighted according to the efficiency-corrected background-subtracted $M_{K\eta'}$ 
distributions in data. The means of these weighted efficiencies are listed in the $\varepsilon$ column of Table \ref{bf}. 

%These efficiencies are strongly dependent on the distribution of $M_{K\eta'}$. While the  
%\Ketapg MC is modelled with a flat $M_{K\eta'}$ distribution between $1.52\,\giga\electronvolt\!/c^2$ and $2.7\,\giga\electronvolt\!/c^2$, the data distribution is not 
%found to agree with this assumption. To correct for this, the 
%efficiencies in 10 bins across $M_{K\eta'}$ are weighted according to the background-subtracted $M_{K\eta'}$ distributions in data and the 
%averages are taken as the calibrated efficiencies shown in Table \ref{bf}. 

The BFs are calculated from the signal yields, calibrated efficiencies, daughter BF products, and the number 
of $B$ mesons in the data sample. Equal production of charged and neutral $B$ meson pairs is assumed. We find 
${\cal B}(B^{+} \rightarrow K^{+} \eta' \gamma) = (3.6\pm1.2\pm0.4) \times 10^{-6}$ and 
${\cal B}(B^{0} \rightarrow K^{0} \eta' \gamma) = (2.5^{+2.4 +0.4}_{-1.9 -0.5}) \times 10^{-6}$, where the first errors are statistical and the second 
systematic. The 90\% CL ULs are found to 
be ${\cal B}(B^{+} \rightarrow K^{+} \eta' \gamma) < 5.6 \times 10^{-6}$, and ${\cal B}(B^{0} \rightarrow K^{0} \eta' \gamma) < 6.4 \times 10^{-6}$. The ULs are 
calculated by integrating the likelihood function with systematic errors included in the physically allowed 
BF region. The UL is then defined as the BF below which 90\% of the integrated likelihood lies. The 
BFs and ULs are measured within the reduced phase-space region $M_{K\eta'}<3.4\,\giga\electronvolt\!/c^2$.

The systematic uncertainties for the charged (neutral) BF include errors on the following processes:
photon detection (2.8\% (2.8\%)); 
$\pi^{0}$ reconstruction (0.8\% (0.5\%)); $K^{0}_{S}$ reconstruction (0.0\% (4.5\%)); $\eta$ reconstruction (3.4\% (3.6\%)); 
charged track detection (3.8\% (5.0\%)); $K^{+}/\pi^{+}$ differentiation (1.4\% (1.5\%)); 
and the calculated number of $B \kern 0.18em\overline{\kern -0.18em B}$ pairs in the data sample (1.4\% (1.4\%)). The statistical uncertainty on the 
MC efficiency 
after calibration is 1.7\% (1.9\%). The data/MC LR selection efficiency difference is estimated to be 3.7\%(3.7\%) using 
a $B \rightarrow K^{*}(892) \gamma$ control sample. 
%results using a $B \rightarrow K^{*}(892) \gamma$ control sample. The uncertainty stemming from the difference in LR cut efficiency 
%between MC and data is estimated to be 3.7\% for both results using a $B \rightarrow K^{*}(892) \gamma$ control sample. 
The vetoes on $M_{\eta'\gamma}$ around the $J/\psi$ invariant mass and on $M_{K^{-}\pi^{+}}$ around the 
$D^{0}$ mass are found to affect the reconstruction efficiency by 0.2\% (0.4\%)) and 0.5\% (0.0\%) respectively. 
The bias study results have an uncertainty of 1.5\% (3.5\%).
The correction of the signal reconstruction efficiency from the $M_{K\eta'}$ distributions has an uncertainty of 1.2\% (1.5\%). 
The effect of $K^{0}_{S} \eta' \gamma$ events being reconstructed as $K^{+} \eta' \gamma$ and entering the incorrect fit distribution, and 
{\it vice versa}, gives a  $-6\% (+6\%)$ efficiency uncertainty.
%a negative 6\% uncertainty on the $K^{+} \eta' \gamma$ reconstruction efficiency, and 
%a corresponding positive 6\% uncertainty for the $K^{0}_{S} \eta' \gamma$ reconstruction efficiency. 
All fixed parameters in the fit to data are varied by $\pm1\sigma$ (the $b \rightarrow c$ and $b \rightarrow u,d,s$ yields 
by $3\sigma$), yielding uncertainties of ($+6.5\%, -6.7\%$) for $B^{+} \rightarrow K^{+} \eta' \gamma$, and ($+11.7\%, -16.6\%$) for 
$B^{0} \rightarrow K^{0}_{S} \eta' \gamma$.
%; the effects on the data yield are added in quadrature and taken as following systematic errors: 
 
The existing measurements of $B \rightarrow X_{s} \gamma$ and $B \rightarrow X_{s} l^{+} l^{-}$ both rely heavily on the accuracy of the $X_{s}$ 
hadronization model performed by 
the JETSET \cite{jetset} program \cite{xsll}\cite{nishACP}\cite{babarACP}. A large sample of inclusive $B \rightarrow X_{s} \gamma$ MC was generated 
with the $X_{s}$ system hadronized by JETSET according to the Kagan-Neubert model \cite{KN} and the mass of the $b$ quark set to 
$4.75\,\giga\electronvolt\!/c^2$.
From this sample, the model predicts the BF of $B^{+} \rightarrow K^{+} \eta' \gamma$ events in the region $M_{K\eta'}<3.4\,\giga\electronvolt\!/c^2$ 
to be $(1.8\pm0.2)\times10^{-6}$, while we measure $(3.6\pm1.2\pm0.4) \times 10^{-6}$. 
In addition, this model predicts ${\cal B}(B \rightarrow K \eta \gamma)$ to be $(8.2\pm0.9)\times10^{-6}$ using the 
same sample, which can be compared to the measured BF of $(7.9\pm0.9)\times10^{-6}$ \cite{PDG}. 
With the current statistics, the BFs for both $B^{+} \rightarrow K^{+} \eta' \gamma$  and $B \rightarrow K \eta \gamma$ 
are consistent with the Kagan-Neubert model.
%From this sample, the BF of $B^{+} \rightarrow K^{+} \eta' \gamma$ events in the region $M_{K\eta'}<3.4\,\giga\electronvolt\!/c^2$ was calculated to be $(1.8\pm0.2)\times10^{-6}$. 
%${\cal B}(B \rightarrow K \eta \gamma)$ was measured to be $(8.2\pm0.9)\times10^{-6}$ using the 
%same sample, which is in agreement with the measured BF of $(7.9\pm0.9)\times10^{-6}$ \cite{PDG}. 
%JETSET is producing the BF heirarcy of these 
%decays correctly but the magnitude of ${\cal BF}($B \rightarrow K \eta' \gamma$)$ is too low.
%This result is in agreement with our measured BF, and JETSET is shown to correctly model $B^{+} \rightarrow K^{+} \eta' \gamma$ 
%with the above parameters.

%A large sample of inclusive $B \rightarrow X_{s} \gamma$ MC was generated with the $X_{s}$ hadronized by JETSET \cite{jetset}. The $M_{X_{s}}$ spectrum 
%was reweighted to match the prediction of the Kagan-Neubert model \cite{KN} with the mass of the $b$ quark ($m_{b}$) set to 
%$4.75\,\giga\electronvolt\!/c^2$. From this sample, we calculate the branching fraction of $B^{+} \rightarrow K^{+} \eta' \gamma$ events in the region $M_{K\eta'}<3.4\,\giga\electronvolt\!/c^2$ 
%to be modelled as $(3.6\pm0.3)\times10^{-6}$. This is in agreement with our measured experimental branching fraction.

In conclusion, we report the first evidence of the decay $B^{+} \rightarrow K^{+} \eta' \gamma$ with a partial BF of 
${\cal B}(B^{+} \rightarrow K^{+} \eta' \gamma) = (3.6\pm1.2\pm0.4) \times 10^{-6}$ and a 
significance of $3.3\sigma$ in the region $M_{K\eta'}<3.4\,\giga\electronvolt\!/c^2$. 
We also set a 90\% CL UL of ${\cal B}(B^{0} \rightarrow K^{0} \eta' \gamma) < 6.4 \times 10^{-6}$ in the same $M_{K\eta'}$ region.

We thank the KEKB group for excellent operation of the
accelerator, the KEK cryogenics group for efficient solenoid
operations, and the KEK computer group and
the NII for valuable computing and SINET3 network support.  
We acknowledge support from MEXT, JSPS and Nagoya's TLPRC (Japan);
ARC and DIISR (Australia); NSFC (China); 
DST (India); MOEHRD and KOSEF (Korea); MNiSW (Poland); 
MES and RFAAE (Russia); ARRS (Slovenia); SNSF (Switzerland); 
NSC and MOE (Taiwan); and DOE (USA).

\begin{comment}
We thank the KEKB group for the excellent operation of the
accelerator, the KEK cryogenics group for the efficient
operation of the solenoid, and the KEK computer group and
the National Institute of Informatics for valuable computing
and SINET3 network support.  We acknowledge support from
the Ministry of Education, Culture, Sports, Science, and
Technology (MEXT) of Japan, the Japan Society for the 
Promotion of Science (JSPS), and the Tau-Lepton Physics 
Research Center of Nagoya University; 
the Australian Research Council and the Australian 
Department of Industry, Innovation, Science and Research;
the National Natural Science Foundation of China under
contract No.~10575109, 10775142, 10875115 and 10825524; 
the Department of Science and Technology of India; 
the BK21 program of the Ministry of Education of Korea, 
the CHEP src program and Basic Research program (grant 
No. R01-2008-000-10477-0) of the 
Korea Science and Engineering Foundation;
the Polish Ministry of Science and Higher Education;
the Ministry of Education and Science of the Russian
Federation and the Russian Federal Agency for Atomic Energy;
the Slovenian Research Agency;  the Swiss
National Science Foundation; the National Science Council
and the Ministry of Education of Taiwan; and the U.S.\
Department of Energy.
This work is supported by a Grant-in-Aid from MEXT for 
Science Research in a Priority Area ("New Development of 
Flavor Physics"), and from JSPS for Creative Scientific 
Research ("Evolution of Tau-lepton Physics").
\end{comment}


\begin{thebibliography}{99}
\bibitem{CC}
The charge-conjugate modes of all decays in this paper are implied unless explicitly stated otherwise.
\end{thebibliography}
\end{document}